\documentclass[12pt,epsf]{article}
\usepackage{epsfig}
\usepackage{amssymb}  
\usepackage{amsthm}
\usepackage{amsmath}
\let\eqref\cref
\newcommand{\onefigure}[2]{\begin{figure}[htbp]
\begin{center}\leavevmode\epsfbox{#1.eps}\end{center}\caption{#2\label{#1}}
\end{figure}}

\renewcommand{\thanks}[1]{\footnote{#1}} 

\newcommand{\be}{\begin{equation}}
\newcommand{\ee}{\end{equation}}
\newcommand{\bea}{\begin{eqnarray}}
\newcommand{\eea}{\end{eqnarray}}
\newcommand{\half}{{1\over 2}}

\def\stacksymbols#1#2#3#4{\def\theguybelow{#2}
\def\verticalposition{\lower#3pt}
\def\spacingwithinsymbol{\baselineskip0pt\lineskip#4pt}
\mathrel{\mathpalette\intermediary#1}}
\def\intermediary#1#2{\verticalposition\vbox{\spacingwithinsymbol
\everycr={}\tabskip0pt
\halign{$\mathsurround0pt#1\hfil##\hfil$\crcr#2\crcr
\theguybelow\crcr}}}

\begin{document}

\pagestyle{empty}

\bigskip\bigskip
\begin{center}
{\bf \large A Plausible Path Towards Unification of Interactions via Gauge Fields Consistent with the Equivalence Principle-II}
\end{center}

\begin{center}
James Lindesay\footnote{E-mail address, jlindesay@howard.edu} \\
Computational Physics Laboratory \\
Howard University,
Washington, D.C. 20059 
\end{center}

\begin{abstract} 

An extension of the Lorentz group that includes
generators $\Gamma^\mu$ carrying a space-time index has been previously demonstrated to 
\emph{explicitly} construct the Minkowski metric \emph{within} the internal group space as a 
consequence of the non-vanishing commutation relations between those generators. Fields that
transform under representations of this extended group can fundamentally incorporate
microscopic causality as a crucial property of physical fields.  The first part of this exploration
focused on the fundamental representation fermions (which satisfy the Dirac equation), and
explored additional internal symmetries associated with those fermions.  Any interactions that could
result from those symmetries were demonstrated to necessarily be consistent with gravitational equivalence
under curvilinear extensions of the (abelian) space-time translations.

The first boson representation of this algebra is the focus of this second paper.  In particular,
the equations of motion for a group of massive vector bosons are degenerate with that of massless
vector bosons, allowing them to unitarily mix to form physical states with differing masses
and dynamics.  Thus, this representation exhibits a potential for enhancing insights into the standard
modeling of electro-weak mixing of bosons.  The various spinors that represent these bosons exhibit kinematic
factors, and those factors are related during unitary mixing to generate the resultant physical states.  For this
reason, analytic and kinematic coincidences associated with known electro-weak masses will be explored
for insights into possible predictive relationships between their masses and those of this first causal
boson representation.  To conclude, a plausible model will be constructed using examined
coincidences for critique and insights into the potential viability of the approach.


\end{abstract}

\bigskip 

\setcounter{equation}{0}
\section{Introduction}
\indent \indent

The relativistic formulation of electron dynamics by Dirac\cite{Dirac} incorporates
the charge-conjugate spinor components necessary to insure microscopic causality\cite{WeinbergQFT}
into an equation that is linear in the energy-momentum operator.
Some properties of fundamental fermions that require
$\hat{\Gamma}^\mu \: \hat{P}_\mu$  to be a Lorentz scalar operation satisfying
\be
\mathbf{\Gamma}^\mu \cdot {\hbar \over i} { \partial \over \partial x^\mu} \,
\hat{\mathbf{\Psi}}_{\gamma}^{(\Gamma)}
(\vec{x}) = -\gamma m  c  \: \hat{\mathbf{\Psi}}_{\gamma}^{(\Gamma)}(\vec{x}) ,
\label{CausalSpinorFieldEqn}
\ee
have been explored in part I of this presentation\cite{PlausibleI}. 
In this formulation, the masses (which are always non-negative) label the standard states, and the factor
$\gamma$ is the eigenvalue of the hermitian operator $\hat{\Gamma}^0$.  Dirac fields transform under
the $\Gamma=\half$ finite dimensional representation of this extended group, and the Dirac matrices
 $\gamma^\mu$ are given by $\gamma^\mu=2 \mathbf{\Gamma}^\mu$.

In this paper, spinors transforming under the $\Gamma=1$ finite dimensional representation of this
extended group will be developed.  In particular, the eigenvalues of the hermitian 10$\times$10 matrix 
$\mathbf{\Gamma}^0$ take on integral values $-1 \le \gamma \le +1$, and eigenvalues
of angular momentum satisfy  $0 \le J \le \Gamma$ and $-J \le \gamma \le J$, providing
associative relationships between vector and scalar components.  
A direct examination of (\ref{CausalSpinorFieldEqn}) for this representation expresses a degeneracy
in the equation for $\gamma=0$ massive components, along with massless fields.  For this
reason, a focus will be placed upon the potential of these fields towards enhancing standard modeling
of electro-weak bosons.
Plausible kinematic coincidences relating various electro-weak parameters
using presently known phenomenology will be explored, in hopes of motivating the development of a
compelling model describing general transformations of physical states.  Some
exemplary coincidences that depend upon un-discovered dark particle states and
have been presented elsewhere\cite{JLtoIJTP2022} will be expanded and
critiqued in light of more recent experimental measurements.  To end the discussion,
plausible models that depends \emph{only} on known electro-weak parameters will be developed as
further exemplars of the potential usefulness of this approach.


\setcounter{equation}{0}
\section{A Brief Extended Lorentz Group Primer}
\indent 

To begin, a brief summary of the general properties of the extended Lorentz group will
be developed.  A more detailed
exposition can be found in \cite{PlausibleI} or \cite{JLFQG}.

\subsection{Group algebra}

The group algebra inclusive of operators $\Gamma^\mu$ requires the non-vanishing commutators to
satisfy
\be
\begin{array}{lll}
\left [ \Gamma^0 \, , \, \Gamma^k \right] \: = \: {i \over \hbar} \, K_k  , &
\left [ \Gamma^0 \, , \, J_k \right] \: = \: 0  , &
\left [ \Gamma^0 \, , \, K_k \right] \: = \: -i  \hbar \,  \Gamma^k  , \\
\left [ \Gamma^j \, , \, \Gamma^k \right] \: = \: -{ i \over \hbar} \, \epsilon_{j k m} \, J_m  , &
\left [ \Gamma^j \, , \, J_k \right] \: = \: i \hbar \, \epsilon_{j k m} \, \Gamma^m  , &
\left [ \Gamma^j \, , \, K_k \right] \: = \: -i \hbar \, \delta_{j k} \, \Gamma^0  .
\end{array}
\label{ExtLorentzGroupEqns}
\ee
Operators $J_k$ and $\Gamma^0$ are hermitian observables, and $K_k$ and $\Gamma^k$ are anti-hermitian. 
For general algebras satisfying
$\left [ \hat{G}_r \, , \, \hat{G}_s \right ] \: = \: -i \, \sum_m \left ( c_s \right ) _r ^m \, \hat{G}_m $,
the structure constants can be used to develop a group metric
$\eta_{a b} \: \equiv \: \sum_{s \, r} \left ( c_a \right )_r ^s \, \left ( c_b \right )_s ^r, \:
\eta^{ab}\equiv  ((\eta)^{-1})_{ab}$ that can construct the Casimir operator
$\hat{C}_\mathcal{G} \equiv \sum_{r s} \eta^{G_r G_s} \hat{G}_r \hat{G}_s$ whose
eigenvalues label the irreducible representations.
The Casimir operator for this extended Lorentz group is 
$C_\Gamma \: = \: {1 \over 6} \left [ \,  ( \underline{J} \cdot \underline{J} \,-\, \underline{K} \cdot \underline{K} )/\hbar^2
\,+\, \Gamma^0 \, \Gamma^0 \,-\, \underline{\Gamma} \cdot \underline{\Gamma} \right ]$, and finite dimensional
representations have dimension $N_\Gamma \: = \: {1 \over 3} (\Gamma + 1) (2 \Gamma + 1) (2 \Gamma + 3)=10$
for the $\Gamma=1$ representation being discussed.  Furthermore, a metric for operators carrying the space time
indexes under Lorentz sub-group transformations takes the form
$\eta^{\Gamma^\mu \, \Gamma^\nu} \: = \: -{1 \over 6} \, \eta_{\mu \, \nu}$ in terms of the Minkowski metric.
\emph{General} spinor forms that satisfy (\ref{ExtLorentzGroupEqns}) are developed in reference \cite{JLFQG}.

\subsection{Kinematics of spinors}

Dimensionless spinors $\mathbf{u}_{\gamma}^{(\Gamma)}$ generally satisfy the equation
\be
\mathbf{\Gamma}^\mu p_\mu \mathbf{u}_{\gamma}^{(\Gamma)}(\underline{\beta},s,s_z) = 
-\gamma \, m_{(st)} c \, \mathbf{u}_{\gamma}^{(\Gamma)}(\underline{\beta},s,s_z),
\label{SpinorEquation}
\ee
where $\underline{\beta}={\underline{p}/ \epsilon}$ defines the relative Lorentz frame of the state for massive particles.
The standard state is typically identified with $\underline{\beta}=0$. 
For massless particles, the standard state is traditionally specified in terms of z-moving helicity
states with contra-variant 4-momentum components $\vec{p}_{(st)}$=(1,0,0,1).  More generally,
standard state massless 4-momentum can be associate with $\vec{p}_{(st)}=\mu_{(st)}c$(1,0,0,1), where
$\mu_{(st)}$ must be a definable kinematic scale uniquely associated with the creation of that state.

The definition of the Dirac adjoint spinor $\bar{\mathbf{u}}\equiv \mathbf{u}^\dagger \mathbf{g}_\Gamma$
to be identified with Dirac products between spinors requires the development of a $N_\Gamma$ dimensional
matrix that satisfies ${\mathbf{G}_s}^\dagger = \mathbf{g}_\Gamma \mathbf{G}_s \mathbf{g}_\Gamma$ for
all finite dimensional forms of the generators of the extended group.  This matrix turns out to be purely
diagonal with elements $\mathbf{g}_\Gamma=(-1)^{\Gamma-\gamma}$, and only corresponds to
$\mathbf{g}_\half=\gamma^0$ for Dirac fermions.



\setcounter{equation}{0}
\section{Causal $\Gamma=1$ Boson States}

\indent

For massive particles, the momentum-space forms of the $\Gamma=1$ spinors depend on
the Lorentz boost $\underline{\beta}$ from the standard states, and the $m \rightarrow 0$
form of \emph{z-moving} spinors are pure numbers, just as was the case for the $\Gamma=\half$ representation spinors. 
\emph{Unlike} the $\Gamma=\half$ fields, the form (\ref{CausalSpinorFieldEqn}) includes a degenerate
set of $\gamma=0$ distinct states that can be mixed while continuing to satisfy the field equation. 
Unitary mixing of these degenerate states establishes relationships
between the kinematic spinors in terms of invariant mixing angles and mass ratios.

Of particular interest,
standard model electro-weak bosons include electrically charged massive $W^\pm$ bosons and
a massive neutral scalar $H$ boson, as well as the massive electrically neutral vector $Z$ boson
and photon $A$ that result from the
unitary mixing of the neutral $W^3$ and $B$ bosons\cite{Weinberg,Peskin}. 
In a prior publication\cite{JLtoIJTP2022}, kinematic conditions resulting from
the unitary mixing of the degenerate $\Gamma=1$ bosonic causal spinors  were demonstrated
to potentially model electro-weak bosons consistent with
experimental results.  The matrix representations of the generators are 
10-dimensional, inherently relating scalar and vector components.
In what follows, plausible relationships and kinematic coincidences involving electro-weak and electromagnetically
dark bosons will be further critiqued and explored. 
Henceforth, natural units with $\hbar\rightarrow1, c\rightarrow 1$ will be utilized,
and Einstein's summation convention over repeated \textit{super}/\textit{sub}scripts will be assumed.


\subsection{General form and properties of bosonic spinors}

The $\Gamma=1$ momentum-space spinor fields $\mathbf{\Phi}$ of the fields described in (\ref{CausalSpinorFieldEqn})
satisfy (\ref{SpinorEquation}) with $\Gamma=1$.  The spin values $s$
are integers satisfying $0 \le s \le 1$, and the eigenvalues of $\mathbf{\Gamma}^0$ satisfying
$-s \le \gamma \le +s$ represent an additional set of discrete integer valued quantum numbers. 
The $\mathbf{\Gamma}^\beta$ are 10$\times$10 matrices\cite{JLFQG10x10} that will be
ordered with the first component being  the single scalar component with $\gamma=0$,
the next three components representing the $s_z$ eigenstates
with $\gamma=+1$, the following three components representing the $s_z$ eigenstates
with $\gamma=0$, and the final three components representing the $s_z$ eigenstates
with $\gamma=-1$. The representation is irreducible due to the mixing of components
in the anti-hermitian matrices $\mathbf{\Gamma}^j$ and $\mathbf{K}_j$.

All Dirac orthonormalized z-moving spinors $\mathbf{u}(\beta_z,s,s_z)$ satisfying
$\mathbf{u}_a^\dagger \mathbf{g}_{\Gamma=1} \mathbf{u}_b  \equiv \bar{\mathbf{u}}_a \mathbf{u}_b=
(-1)^{1-\gamma_b}\delta_{a b}$
have components that become singular when $m\rightarrow 0$, where $\mathbf{g}_{\Gamma=1}$ is a diagonal
matrix with elements $(-1)^{1-\gamma}$ for all spin states. 
Other than the subset of $\gamma=0$ spinors, the \emph{hermitian} normalized spinors are not orthogonal. 
However, only the hermitian orthonormalized $\gamma=0$ spinors are non-singular for massless systems.  Massive
standard state (i.e. at rest) spinors are always eigenstates of
$\mathbf{\Gamma}^0$ satisfying $\left .\mathbf{\Gamma}^0 \mathbf{u}_a =
\gamma_a \: \mathbf{u}_a \right |_{\beta \rightarrow 0}$., where $a=1$ for the $\gamma=0$ scalar spinor,
for the $\gamma=+1$ vectors $a \in \{2,3,4\}$, for the degenerate $\gamma=0$ vectors $a \in \{5,6,7\}$,
and for the $\gamma=-1$ (anti-particle) vectors $a \in \{8,9,10\}$.

The massless spinors are \emph{never} eigenstates of $\mathbf{\Gamma}^0$.
However, they \emph{do} satisfy $p_\mu \mathbf{\Gamma}^\mu \mathbf{u}_a |_{m=0}= 0$.
Generally, for massless spinors the forms
$\mathbf{\mathcal{A}}_{\gamma,s_z}^{(s) \: \beta} (\vec{p})\equiv \mathbf{\Gamma}^\beta \, 
\mathbf{\Phi}_{\gamma,s_z}^{(1,s)}(\vec{p})$ satisfying (\ref{SpinorEquation}) are
momentum space spinor forms transverse to the 4-momentum
$p_\beta \mathbf{\mathcal{A}}_{\gamma,s_z}^{(s) \: \beta} (\vec{p})=0$. 
The configuration space form of these fields thus satisfy the Lorentz gauge.

Only the components $\mathbf{\mathcal{A}}_{0,0}^{(s=0)\beta} (\vec{p})$ generate \emph{new} spinor forms
that are also eigenstates of $\mathbf{S}_z \: {  \mathbf{\mathcal{A}}_{0,0}^{(s=0) x}
\pm i \mathbf{\mathcal{A}}_{0,0}^{(s=0) y} \over \sqrt{2} } =\pm 1\:  {  \mathbf{\mathcal{A}}_{0,0}^{(s=0) x}
\pm i \mathbf{\mathcal{A}}_{0,0}^{(s=0) y} \over \sqrt{2} }$,
$\mathbf{S}^2 \: {  \mathbf{\mathcal{A}}_{0,0}^{(s=0) x}
\pm i \mathbf{\mathcal{A}}_{0,0}^{(s=0) y} \over \sqrt{2} } =2\:  {  \mathbf{\mathcal{A}}_{0,0}^{(s=0) x}
\pm i \mathbf{\mathcal{A}}_{0,0}^{(s=0) y} \over \sqrt{2} }$,
\emph{and} can also orthogonally mix the original spinors $\mathbf{\Phi}_{\gamma,s_z}^{(1,s)}(\vec{p})$. 
The forms of $\mathbf{\mathcal{A}}_{0,0}^{(s=1)\beta} (\vec{p})$
are \emph{not} spinor eigenstates of angular momentum $\mathbf{S}_z,\mathbf{S}^2$. 
Thus, the configuration space form of the $\mathbf{\mathcal{A}}_{0,0}^{(0) \beta}$ fields can represent
covariant gauge potentials that directly generate the field strengths $F_{\mu \nu}\equiv
\partial_\mu \mathbf{\mathcal{A}}_{0,0 \: \nu}^{(0)}-\partial_\nu \mathbf{\mathcal{A}}_{0,0 \: \mu}^{(0)}$
that express common representations of spatial polarizations.


\subsection{Unitary kinematic mixing of degenerate components}
\indent 

The mixing of previously degenerate eigenstates to develop dynamically alternative representations
is a phenomenon that sometimes occurs in perturbative atomic physics (e.g. Stark effect). 
In this subsection, relationships for unitary kinematic
mixing of degenerate $(\Gamma=1,\gamma=0)$ states will be briefly developed. 
The focus will be on mixing degenerate spinors moving parallel to the z-axis expressed
in terms of a kinematic parameter $\zeta(\beta_z)$ which is a function only of the Lorentz transformation from
the standard state $\beta_z$.  The \emph{hermitian} normalized scalar spinor has the form 
\be
\mathbf{U}_X (\zeta_X,s=0,s_z=0)=\left (
\begin{array}{c}
\cos \zeta_X \\ 0 \\ -{\sin \zeta_X \over \sqrt{2}}  \\ 0 \\ 0 \\ 0 \\
 0 \\ 0 \\ {\sin \zeta_X \over \sqrt{2}} \\ 0
\end{array} \right )
\stacksymbols{\Longrightarrow}{{}_{m_X \rightarrow 0}}{4}{0.1}
\left (
\begin{array}{c}
{1 \over \sqrt{2}} \\ 0 \\-{1 \over 2}  \\ 0 \\ 0 \\
0 \\ 0 \\ 0 \\ {1 \over2} \\ 0
\end{array} \right ),
\ee
where the dimensionless \emph{kinematic angle} for particle X is defined by
\be
\begin{array}{c}
\zeta_X \equiv \sin^{-1}\left (
p_X \over \sqrt{m_X^2 + 2 p_X^2} \right ) =
  \sin^{-1}\sqrt{\epsilon_X^2 -m_X^2 \over 2 \epsilon_X^2 -m_X^2}\quad\\
\quad \rightarrow
p_X = m_X {\sin \zeta_X \over \sqrt{\cos 2 \zeta_X}},  \:  
\epsilon_X = m_X {\cos \zeta_X \over \sqrt{\cos 2 \zeta_X}},
\end{array}
\label{zetamEqn}
\ee
with the Lorentz transformation from the standard state parameterized by $\beta_z=\tan \zeta_X$.
It is noteworthy that systems with identical ${p \over \epsilon}$ are co-moving, and thus have
identical spinor forms.  Massive spinors have kinematic angles in the range $-{\pi \over 4}<\zeta_X
<{\pi \over 4}$, and massless spinors have $\zeta_X=\pm {\pi \over 4}$.
The connection between $\gamma=0$ Dirac orthonormal spinors $\mathbf{u}_X$ discussed previously
and hermitian normalized spinors $\mathbf{U}_X$
is given by $\mathbf{u}_X={\mathbf{U}_X \over \sqrt{\cos 2 \zeta_X}}$, which clearly demonstrates
a breakdown of Dirac-type normalization for \emph{massless} bosons.

Consider the general mixing of orthogonal states $X+V\Rightarrow R+Y$.  Unitary mixing of the spinor components
requires
\be
\begin{array}{r}
\cos \theta_{XV} \: \mathbf{U}_{X} ( \zeta_X,s,s_z) +
\sin \theta_{XV} \: \mathbf{U}_{V}( \zeta_V,s,s_z)  =
\mathbf{U}_{R} ( \zeta_{R},s,s_z)  , \\ 
-\sin\theta_{XV} \: \mathbf{U}_{X} ( \zeta_X',s,s_z) +
\cos \theta_{XV} \: \mathbf{U}_{V} ( \zeta_V',s,s_z)  =
\mathbf{U}_{Y} ( \zeta_{Y},s,s_z) ,
\end{array}
\label{XVmixingEqn}
\ee
where $R$ and $Y$ types interchange for $\theta_{XV}\rightarrow -\theta_{XV}$.
For instance, if $\zeta_X >0$ the spinor $\mathbf{U}_{R} ( \zeta_{R},s,s_z)$ has the form
$\mathbf{U}_{X} ( \zeta_X-\theta_{XV} ,s,s_z)$, and the spinor $\mathbf{U}_{Y} ( \zeta_{Y},s,s_z)$
has the form $\mathbf{U}_{V} ( \zeta_X-\theta_{XV} ,s,s_z)$. 
Orthogonality requires that if $X$ is mass/energy-like, then $V$ must be tachyon/momentum-like,
guaranteed by kinematic angles satisfying $\zeta_V=\zeta_X-{\pi \over 2}$.

Expressing the tachyonic form for system $V$ via $m_V \rightarrow i \mu_V$,
orthogonal mixing relates the kinematic angles as follows:
\be
p_V \rightarrow \mp {\mu_V \over m_X} \sqrt{m_X^2 + p_X^2} \quad , \quad
\sqrt{m_V^2 + p_V^2} \rightarrow \pm {\mu_V \over m_X} p_X \quad , \quad
\zeta_V=\zeta_X \mp {\pi \over 2}.
\label{pVEVeqns}
\ee
The resulting spinors through trigonometric re-combination have kinematic
angle $\zeta_R = \zeta_X \mp \theta_{XV}$.  Furthermore, the spinor $\mathbf{u}_{R}$ will have the same
kinematic form as $\mathbf{u}_{X}$, and $\mathbf{u}_{Y}$ will share the orthogonal form of $\mathbf{u}_{V}$. 
For massless resultant spinors, $\mathbf{u}_{R}(\zeta=\pm{\pi \over 4})=\mp \mathbf{u}_{Y}(\zeta=\mp{\pi \over 4})$. 
In most of what follows, only the upper signs will be utilized unless the sign is explicitly demonstrated. 
The resultant energy-momentum of the $\mathbf{R}$ or $\mathbf{Y}$ spinor is required to satisfy
$\vec{P}_X + \vec{P}_V = \vec{P}_{R,Y}$ as appropriate for the resultant mass. 
Given this expression, one can derive the form
of the kinematic angle associated with mixing a massive $X$ with an orthogonal $V$ resulting
in a product with invariant energy $M$, given by
\be
\zeta X ^M _{\mu_V}= \pm \sin^{-1} \left (
\sqrt{  \sqrt{M^4+2 M^2 (\mu_V^2-m_X^2 ) +(\mu_V^2 +m_X^2)^2}  -2 m_X \mu_V   } \over
\sqrt{2} \left ( M^4+2 M^2 (\mu_V^2-m_X^2 ) +(\mu_V^2 +m_X^2)^2   \right )^{1 \over 4}
\right ),
\label{zetaXmixEqn}
\ee
where the sign follows that of the momentum $p_X$.  This form vanishes when $M^2=m_X^2-\mu_V^2$. 
For `bootstrap' self mixing, the kinematic angle takes the value
${\zeta X}_{\mu_X}^{m_X}=\sin^{-1}\sqrt{\half-{1 \over \sqrt{5}}}$. 
Generally, the form $\vec{p} \cdot \vec{x}=m{z \sin \zeta - ct |\cos \zeta| \over \sqrt{\cos (2 \zeta)}}$
that appears in the 4-momentum conserving exponentials of quantum fields differ \emph{only} in the masses of
 co-moving systems.  It should be emphasized that all massive co-moving states will share the same kinematic angle.

Consistency of (\ref{zetaXmixEqn}) with mixing to a \emph{massless} state
$M_{Y}=0 \: \rightarrow \: \zeta_Y=\pm {\pi \over 4}$
requires that $\theta_{XV}=\tan^{-1}{\mu_V \over m_X}$ and $|\zeta X ^0 _{\mu_V}|={\pi \over 4}-\theta_{XV}=
|\zeta X ^{M_{Yxv}} _{\mu_V}|$, where an additional invariant energy $M_{Yxv}$ satisfies this relationship if
$m_X>\mu_V$.  The
energy of this resultant state and an invariant representation label $M_{Yxv}$ satisfy
\be
\epsilon_{Yxv}={m_X^2+\mu_V^2 \over 2 \sqrt{m_X \mu_V}} \: \textnormal{ and } \:
M_{Yxv}=\sqrt{2}\sqrt{m_X^2-\mu_V^2}.
\label{epsilonYeqn}
\ee
Note that generally, $\tan^{-1}{m_V \over m_X}+\tan^{-1}{m_X \over m_V}={\pi \over 2}$,
which relates the mixing angles resulting from interchange of $X$ and $V$.
The kinematic angle of the spinor {\bf{R}} resulting from this mixing thus satisfies
$\zeta_R=\zeta X^M_{\mu_V}-\theta_{XV}=\zeta X^M_{\mu_V}-\tan^{-1}{\mu_V \over m_X}$.
Furthermore, orthogonal \emph{self}-mixing to $M_{XX}=0$ states with ${\zeta X}_{\mu_X}^0=0$
(i.e. $X$ at rest) generates massless states of momenta $\pm m_X$.


\subsection{Modeling of electro-weak bosons}


\subsubsection{Standard electro-weak modeling\label{Sec:EWreview}}

In the standard model of electro-weak interactions, the Lagrangian density
\be
\mathcal{L}_{EW}=\left( (\partial_\mu -i g_W W_\mu^{(a)} \hat{\tau}_a -i g_B B_\mu \hat{Y}) \phi \right )^\dagger \eta^{\mu \nu}
( \partial_\nu -i g_W W_\nu^{(b)} \hat{\tau}_b -i g_B B_\nu \hat{Y}  ) \phi - V(\phi^\dagger \phi)
\label{EWLagrangian}
\ee
manifests local gauge invariance under SU[2]$\times$U[1] transformations, where the $W_\nu^{(b)}$ potentials are associated with
the SU[2] generators $\hat{\tau}_b$, and the $B_\nu$ potentials are associated with a U[1] hypercharge generator $\hat{Y}$.  For a
potential that takes the form $V( {\phi}^\dagger \phi) \cong m_\phi^2 \, \phi^\dagger \phi + \lambda (\phi^\dagger \phi)^2$,
there is a minimum when $\langle\phi \rangle=0$, and \emph{this} ground state maintains the symmetry of the Lagrangian..  

However, if the mass term becomes `tachyonic' $m_\phi \rightarrow \pm i \mu_\phi$ , then
the potential is minimum for \emph{any} particular value of the field that satisfies 
$<\phi^\dagger \phi > \cong {\mu_\phi^2 \over 2 \lambda}$, so that the system settles into a particular ground state
that breaks the symmetry of the Lagrangian. Choosing the $\hat{\phi}^0$ component to manifest the broken symmetry,
longitudinal excitations about this new minimum $\hat{\phi}^0\equiv \langle \phi \rangle+\hat{h}$ then
define a massive Higgs scalar field $\hat{h}$, with real mass $m_H=\sqrt{2}\, \mu_\phi$ and vanishing ground state expectation value.
Furthermore, transverse excitations along the symmetry of the Lagrangian result in massless (Goldstone) modes.

Substitution of the form of  $\hat{\phi}^0$ into (\ref{EWLagrangian}) yields quadratic terms generating
a mass for the gauge bosons $W_\mu^{(\pm )}$ of  $m_{W^{(\pm)}}=\left | {g_W\langle \phi \rangle \over \sqrt{2}}\right |$,
which is independent of the units of charge utilized.
Furthermore, the remaining gauge bosons  $W_\mu^{(3)}$ and $B_\mu$ are mixed to generate
a massive charge-neutral $Z$ boson and a massless photon that appropriately
couples to electric charges of the $W_\mu^{(\pm )}$ bosons using
\be
Z_\mu =W_\mu^{(3)} \cos \vartheta_{w} + B_\mu \sin \vartheta_{w}, \quad
A_\mu= -W_\mu^{(3)} \sin \vartheta_{w} + B_\mu \cos \vartheta_{w}.
\label{WBtoZAstandard}
\ee
The charge coupling to the $W_\mu^{(\pm )}$ that results upon
substitution of (\ref{WBtoZAstandard}) into (\ref{EWLagrangian}), and the quadratic term
in the field $Z_\mu$, allow the following identifications:
\be
e=g_W\sin\vartheta_{w}=g_B \cos \vartheta_{w}, \quad
\cos \vartheta_{w}={m_W \over m_Z}.
\ee


\subsubsection{Unitary mixings involving \bf{H}+$\phi$ spinors}

An algebra of transformations whose boson representation generates spinors with both scalar and vector components
presents an intriguing characteristic of possible relevance to the standard electro-weak bosons. 
To begin an exploration towards identifying electro-weak bosons as causal spinor fields, examine the orthogonal mixing of
the degenerate components of a spinor inclusive of the scalar $H$ with the tachyonic $\mu_\phi$ that induces the
electro-weak symmetry breaking using (\ref{zetaXmixEqn}).

The mixing of $m_H$ with $\mu_\phi={m_H \over \sqrt{2}}$ exhibits several unique properties.  First,
the unitary mixing angle satisfies $\theta_{H\phi}=\tan^{-1}{m_\phi \over m_H}=\tan^{-1}{1 \over \sqrt{2}}$,
defining a massless state with energy $\epsilon_{Yh\varphi} ={3 m_H \over 2^{7 / 4}}$ generated by kinematic angle
$|\zeta H_{m_\phi}^{M=0}|={\pi \over 4}-\theta_{H\phi}$. 
Furthermore \emph{this} combination, uniquely defines kinematic angles
$|\zeta H_{\mu_\phi}^{m_H}|=|\zeta H_{\mu_\phi}^{M=0}|=\sin^{-1}\sqrt{{1 \over 6}(3-2 \sqrt{2})}$,
which implies opposite momenta $p_H(\zeta H_{\mu_\phi}^{m_H})=-p_H(\zeta H_{\mu_\phi}^{0})$. 
This allows the self-generating mix $H+\phi \rightarrow \tilde{H}$ to an $R$-like $\tilde{H}$
with a mass $m_{\tilde{H}}=m_H$ of momentum
$p_H$, along with $H'+\phi \rightarrow \tilde{G}$  to a \emph{massless} $R$-like $\tilde{G}$
at the same invariant $M=m_H$ of momentum $-p_H$,
where $\tilde{G}$ can be associated with a massless mode generated by an $H'$ with opposite kinematic angle
$-\zeta H_{\mu_\phi}^{m_H}$. 
In addition, the mix $H+\phi \rightarrow Y$ at $M=0$ generates a massless  {\bf{Y}}
spinor with the same energy as the $\tilde{G}$, but orthogonal to it.  Similarly,
$H+\phi \rightarrow Y$ at $M=m_H$ generates a massive $Y$-like spinor with the same kinematics
as the $\tilde{G}$. 
This is apparently a type of bootstrap consistency unique amongst particles whose mass is generated by this type
of symmetry breaking. 
The energy-momentum of the massless modes satisfy
$\vec{P}=({3 m_H \over 2^{7 / 4}},0,0,\mp{3 m_H \over 2^{7 / 4}}  )$
from (\ref{epsilonYeqn}). 

Iit is perhaps noteworthy that $H+\phi$ mixing to result in an $R$-like $\phi$  occurs with the $H$ at rest
$\zeta H^{m_\phi}_{\mu_\phi}=0$.
Analogous relationships are satisfied for mixing of the mass scale
$g_W \langle \phi \rangle$ with a tachyonic $\mu_{W}=\left | {g_W\langle \phi \rangle \over \sqrt{2}}\right |$,
as well as for the mixing of $g_B \langle \phi \rangle$ with the tachyonic mass scale
$\mu_B=\left | {g_B\langle \phi \rangle \over \sqrt{2}}\right |=m_W \tan{\theta_W}$, to be discussed next.
Finally, the orthogonal mixing of two {\bf{H}} spinors generates
\emph{both} orthogonal massless modes with the kinematic angle $\zeta H_{\mu_H}^{M=0}=0$ at rest
and resultant massless modes of 4-momentum $\vec{p}=(m_H,0,0,\pm m_H)$,
since $\theta_{H H}={\pi \over 4}$.


\subsubsection{Mixing of W+B spinors}

For $W+B$ mixing, the unique case of invariant mass 
$M={m_W \over \cos  \theta_{WB}} \equiv m_Z$ \emph{analytically} generates an
opposite helicity state for which
\be
\zeta W^{m_W \over \cos \theta_{WB}}_{m_W \tan \theta_{WB}}\equiv \zeta W^{m_Z}_{\mu_B}
 ={\theta_{WB} \over 2} =-(\zeta W^{m_Z}_{\mu_B}-\theta_{WB}) =-\zeta_{Zwb},
\label{WZantimoving}
\ee
i.e. the standard state frame of the {\bf{W}}$_0$ spinor is equal and opposite that of the resultant {\bf{Z}} spinor. 
This means that the spinor structure of any co-moving \emph{massive} states
with equal and opposite momenta can be donated to the
{\bf{W}}$_0$ and {\bf{Z}} spinors while preserving Lorentz invariance, even if those states are initially massless. 
It also defines the mass scale of the $B$ as $\mu_B=m_W \tan \theta_{WB}=\sqrt{m_Z^2-m_W^2}$.

\subsubsection{Analytic coincidences involving electro-weak mass scales}

Several analytic coincidences beyond those already discussed can be ascertained.  For instance, the ratios of
the mass scales $m_H+\mu_\phi$, $g_W\langle\phi\rangle+\mu_W$, and $g_B \langle\phi\rangle + \mu_B$
are identical, implying relationships defined by the fundamental symmetry breaking of the form
\be
|\zeta \langle g\phi\rangle^{\langle g\phi\rangle}_{\mu_W}|=
|\zeta \langle\tilde{g}\phi \rangle^{\langle\tilde{g}\phi\rangle}_{\mu_B}|=|\zeta H^{m_H}_{\mu_\phi}|=
{\pi \over 4}-\theta_{H\phi} =|\zeta H^{0}_{\mu_\phi}|=
|\zeta \langle\tilde{g}\phi \rangle^0_{\mu_B}|=
|\zeta \langle g\phi\rangle^0_{\mu_W}|,
\ee
where for brevity $\langle g\phi \rangle\equiv g_W \langle\phi\rangle$ and $\langle\tilde{g} \phi\rangle\equiv g_B \langle\phi\rangle$.  Also, due to relationships between energy scales of kinematic angles like (\ref{zetaXmixEqn}),
several mixes from rest generate relevant states, e.g.
\be
 \zeta H ^{m_\phi}_{\mu_\phi}=0, \: \zeta\langle g\phi\rangle^{m_W}_{\mu_W}=0,\textnormal{ and } 
\zeta \langle g \phi \rangle^{M_{Awb}}_{\langle \tilde{g} \phi \rangle}=0.
\ee
Examples of co-moving states mixing to massless resultants include
\be
\zeta W^0_{\mu_H}=\zeta H^0_{\mu_W} \:\textnormal{and} \:
\zeta \langle g \phi \rangle^0_{\langle \tilde{g} \phi \rangle} = \zeta W^0_{\mu_B}.
\ee
Furthermore, the exchange $W\leftrightarrow B$ relates the kinematic angles via
\be
\zeta W^{m_Z}_{\mu_B}+\zeta B^{m_Z}_{\mu_W}={\pi \over 4} \textnormal{ and }
\zeta B^0_{\mu_W}+\theta_{WB}={\pi \over 4}.
\ee
Finally, as $M_{Yh\phi}=m_H$ was the case for $H+\phi$ mixing,
$M_{Y\langle g\phi\rangle w}=\langle g \phi \rangle$ is identically true for $\langle g \phi \rangle+W$ mixing.

\subsubsection{Kinematic coincidences involving electro-weak mass scales}

Kinematic coincidences associated with unitary mixing of degenerate boson representations of causal states
will next be explored as a means to motivate the development of a compelling model incorporating the
known properties of electro-weak bosons.  To this end, various kinematic invariants and energies related to
$W,B,Z,H,\phi,\langle g_W \phi \rangle,\langle g_B \phi \rangle$,
and the  various mixing angles will be explored. 

Beyond masses, additional scales were included for examination.  
For instance, the invariant energy that satisfies $|\zeta W^{M_{Awb}}_{\mu_B}|+\theta_{WB}={\pi \over 4}$
perhaps generating a \emph{massless} state labeled by $M_{Awb}\simeq 96 GeV$
in $W+B$ mixing, the  energy
of the photon $\epsilon_{Awb}\simeq 70.7 GeV$ using (\ref{epsilonYeqn}), as well as the analogous energies
(e.g. $\epsilon_{Yh\phi}\simeq 111.7 GeV$) for other mixings, were examined. 
Also of interest was the observation that the kinematic angle
describing the self-mixing bootstrap $\zeta m_m^m=\sin^{-1}
\sqrt{{1 \over 2} - {1 \over \sqrt{5}} }$ is the product of $W+B$ mixing at the bootstrap
invariant energy $M_{bs}\simeq 245.8 GeV$ satisfying $\zeta W_{\mu_B}^{M_{bs}}-\theta_{WB}\equiv\zeta m_m^m$.
Furthermore, a unique invariant energy $M_*\simeq 246.4 GeV$ relating a previously examined kinematic symmetry
involving $W$ and $Z$ threshold energies in the $H$ rest frame\cite{JLtoIJTP2022} given by
$\zeta_Z (\epsilon={m_H^2 \over m_W})=\zeta_W (\epsilon={m_H^2 \over m_Z}) \equiv 
\zeta W^{M_*}_{\mu_B}$, was included.

Previously published predictions of the masses and ratios will here be updated using the 2022 Particle
Data Group (PDG) tables\cite{PDG2022}, as well as compared to the more recent 2022 re-analysis of
the Collider Detector at Fermilab
(CDF) W boson mass data\cite{CDF2022}.  The reported mass values here utilized are as follows:
$m_Z\simeq 91.1876(21)$, $m_H \simeq 125.25(17)$, $m_W^{PDG}\simeq 80.377(12)$, and
$m_W^{CDF}\simeq 80.4335(94)$.
For future comparisons, the difference between the reported PDG and CDF values of the W mass is
$\Delta m_W \equiv m_W^{CDF}-m_W^{PDG}\simeq 4.71 \, \sigma_W^{PDF}\simeq 6.01 \, \sigma_W^{CDF}$.

The phenomenological coincidences associated directly with  various kinematic angles of mixing follow below
in no particular order.  In addition to attempting to find meaningful coincidences with regards to model building,
an additional purpose of such searches is a determination of just how difficult it is to find coincidences consistent
with present observations:
\be
\begin{array}{l}
KC1: \: \zeta H^{m_H}_{\mu_\phi}-\theta_{H\phi}\simeq -\zeta W^{m_H}_{\mu_H} \rightarrow
\begin{array}{l}
m_W\simeq (m_W-0.077 \sigma_W)_{PDG} \textnormal{ or}\\
m_H\simeq (m_H+0.008 \sigma_H)_{PDG},
\end{array} \\
\quad\quad\quad\quad {m_H \over m_W} \simeq {7+\sqrt{113} \over 8 \sqrt{2}},
\end{array}
\label{KC1}
\ee
i.e., $H+\phi$ mixing at $m_H$ generates anti-moving kinematics with $W+H$ mixing at $m_H$;
\be
KC2: \: \zeta W^{\langle g_B \phi \rangle}_{\mu_H} \simeq {\theta_{HW}\over 2}=
{\pi \over 4}-{\theta_{WH}\over 2} \rightarrow
\begin{array}{l}
m_W \simeq (m_W+1.62 \sigma_W)_{PDG} \textnormal{ or}\\
m_H \simeq (m_H-0.367 \sigma_H)_{PDG},
\end{array}
\label{KC2}
\ee
with an interpretation yet to be ascertained;
\be
\begin{array}{c}
\begin{array}{c}
KC3: \quad \: \zeta_H (p_W(\zeta W ^{m_Z}_{\mu_H})) \simeq \zeta W^{m_Z}_{\mu_B} \\
\quad\quad\quad \textnormal{or } p_W(\zeta W^{m_Z}_{\mu_H}) \simeq  p_H({\theta_{WB} \over 2})
\end{array} \rightarrow
\begin{array}{l}
m_W\simeq (m_W+0.801 \sigma_W)_{PDG} \textnormal{ or}\\
m_H\simeq (m_H-1.05 \sigma_H)_{PDG},
\end{array} \\
{m_W \over m_Z}\simeq{2 m_H^3 \over 2 m_H^3 -2 m_H m_W^2 + m_W
\sqrt{m_H^4 +(m_Z^2-m_W^2)^2+2 m_H^2 (m_Z^2+m_W^2)}}
\end{array}
\label{KC3}
\ee
i.e., the kinematic angle of an {\bf{H}} with the momentum of the {\bf{W}} which mixes
with an orthogonal {\bf{H}} to generate a {\bf{Z}} is co-moving with the {\bf{W}} in $W+B$ mixing
to generate a {\bf{Z}}, perhaps suggestive of the $B$ as an off $\mu$-shell version of an $H$;
\be
KC4: \: M_{bs}\simeq M_* \rightarrow
\begin{array}{l}
m_W\simeq (m_W-1.1 \sigma_W)_{PDG} \textnormal{ or}\\
m_H\simeq (m_H-0.751 \sigma_H)_{PDG},
\end{array}
\label{KC4}
\ee
i.e., the mass of the bootstrap generated by $W+B$ mixing is same as the mass generated by the $W+B$
energy invariant symmetric under $m_W \leftrightarrow m_Z$ exchange;
\be
KC5: \: \zeta Z^{M_*}_{\mu_\phi}\simeq \zeta_W (\epsilon_W=m_H) \rightarrow
\begin{array}{l}
m_W\simeq (m_W+0.219 \sigma_W)_{CDF} \textnormal{ or}\\
m_H\simeq (m_H-0.031 \sigma_H)_{CDF},
\end{array}
\label{KC5}
\ee
i.e., the invariant energy $M_*$ for $W+B$ mixing from (\ref{KC4}) is the same as that for $Z+\phi$
mixing (the only single coincidence presented that is most consistent with the CDF results);
\be
KC6: \: \zeta H^{M_{Awb}}_{m_W} \simeq 0 \rightarrow
\begin{array}{l}
m_W\simeq (m_W+1.62 \sigma_W)_{PDG} \textnormal{ or}\\
m_H\simeq (m_H-0.367 \sigma_H)_{PDG},
\end{array}
\label{KC6}
\ee
i.e., the $H$ is at rest in $H+W$ mixing at the invariant energy $M_{Awb}$ for (possible) mixing
of $W+B$ to a same helicity massless state;
\be
\begin{array}{c}
KC7: \: \zeta H^{m_H}_{\mu_B}-\theta_{HB}\simeq -{\theta_{WB}\over 2} \rightarrow
\begin{array}{l}
m_W\simeq (m_W-4.86 \sigma_W)_{PDG}\quad { or}\\
m_H\simeq (m_H+0.37 \sigma_H) \textnormal{using } m_W^{PDG} { or}\\
m_H\simeq (m_H+0.72 \sigma_H) \textnormal{using } m_W^{CDF},
\end{array}\\
{m_H \over m_Z}\simeq \half \left (\cos \theta_{WB}+\left ( \cos \theta_{WB} (1-\sin \theta_{WB})^2\right )^{1/3}+
\left ( \cos \theta_{WB} \left (1+\sin \theta_{WB}\right )^2 \right )^{1/3} \right ),
\end{array}
\label{KC7}
\ee
i.e., the spinor generated by $H+B$ mixing to generate a new {\bf{H}}
is co-moving with the {\bf{Z}} generated from $W+B$ mixing, and anti-moving with that {\bf{W}};
\be
\begin{array}{c}
KC11: \: \zeta W^{<g_W \phi>}_{<g_B \phi>}-\theta_{W<g_B \phi>} \simeq -{\theta_{WB}\over 2} \rightarrow
\begin{array}{l}
 m_W\simeq (m_W\simeq+1.3 \sigma_W)_{PDG} \textnormal{ or}\\
m_W\simeq (m_W-4.3 \sigma_H)_{CDF},
\end{array}\\
{m_W \over m_Z}\simeq \sqrt{{2 \over 7}\left (8+5\sqrt{2}-\sqrt{79+52\sqrt{2}}\right )},
\end{array}
\label{KC11}
\ee
 a relationship involving a {\bf{W}} mixing with energy scales in the Lagrangian.

Any consistent combination of two of the previous kinematic coincidences results in a determination
of two mass ratios.  Since the $Z$ mass is presently best measured, the closest predictions of
$m_W$ and $m_H$ resulting from a requirement of two coincidences follow:\\
(\ref{KC1})+(\ref{KC3}) $\rightarrow  m_W\simeq (m_W+0.881 \sigma_W)_{PDG}$ \emph{and}
$m_H\simeq (m_H+0.105 \sigma_H)_{PDG}$,\\
(\ref{KC1})+(\ref{KC4}) $\rightarrow  m_W\simeq (m_W-0.957 \sigma_W)_{PDG}$ \emph{and}
$m_H\simeq (m_H-0.097 \sigma_H)_{PDG}$,\\
(\ref{KC2})+(\ref{KC3}) $\rightarrow  m_W\simeq (m_W+0.631 \sigma_W)_{PDG}$ \emph{and}
$m_H\simeq (m_H-0.224 \sigma_H)_{PDG}$,\\
(\ref{KC2})+(\ref{KC4}) $\rightarrow  m_W\simeq (m_W-0.422 \sigma_W)_{CDF}$ \emph{and}
$m_H\simeq (m_H-0.462 \sigma_H)_{CDF}$,\\
(\ref{KC3})+(\ref{KC4}) $\rightarrow  m_W\simeq (m_W+0.152 \sigma_W)_{PDG}$ \emph{and}
$m_H\simeq (m_H-0.854 \sigma_H)_{PDG}$,\\
(\ref{KC3})+(\ref{KC5}) $\rightarrow  m_W\simeq (m_W+0.159 \sigma_W)_{PDG}$ \emph{and}
$m_H\simeq (m_H-0.846 \sigma_H)_{PDG}$,\\
(\ref{KC3})+(\ref{KC6}) $\rightarrow  m_W\simeq (m_W+0.631 \sigma_W)_{PDG}$ \emph{and}
$m_H\simeq (m_H-0.224 \sigma_H)_{PDG}$,\\
(\ref{KC3})+(\ref{KC7}) $\rightarrow  m_W\simeq (m_W+1.14 \sigma_W)_{PDG}$ \emph{and}
$m_H\simeq (m_H+0.452 \sigma_H)_{PDG}$,\\
(\ref{KC4})+(\ref{KC5}) $\rightarrow  m_W\simeq (m_W+0.144 \sigma_W)_{PDG}$ \emph{and}
$m_H\simeq (m_H-0.848 \sigma_H)_{PDG}$.
The coincidence (\ref{KC7}) has been purposefully included as a cautionary example of an individual
coincidence several standard deviations from \emph{presently}
measured results providing better predictions when combined with another coincidence.

\subsubsection{Kinematic coincidences involving photon normalization}

For clarity, this section will utilize gaussian units for all charges $e_G\equiv \sqrt{\alpha}=g_W \sin \theta_{WB}=
g_B \cos \theta_{WB}$.  Typical normalizations for the photon field utilize either forms of the type
$e_G \sqrt{{\epsilon_s \over \epsilon_A}}$ (see e.g. \cite{WeinbergQFT,JLFQG,Kaku,BjDrell})
or $e_G {\epsilon_s \over \epsilon_A}$ (see e.g. \cite{RedShelf,Srednicki}),
where $\epsilon_A$ is the energy of the photon, and $\epsilon_s$ is its standard state energy,
customarily assigned to unity ($\epsilon_s \rightarrow 1$). 
The reported value of the fine structure constant here utilized is ${1\over \alpha_{PDG}}\simeq 137.03599907$.
Two coincidences presented only constrain $m_H$:
\be
\alpha H1: \: {1 \over \sqrt{2}}  (g_W \sin \theta_{W \phi}) \:\sqrt{\epsilon_{Yh\phi}} \simeq 1
\rightarrow m_H\simeq (m_H-0.79 \sigma_H)_{PDG},
\label{NWA1}
\ee
which defines the energy of the massless resultant of $H+\phi$ mixing as the standard
state energy of electromagnetic quanta, in terms of the coupling defined for $W+\phi$ mixing, and
\be
\alpha H2: \: {\sqrt{4 \pi} e_G \cos \theta_{H\phi} \over \sin \theta_{Z\phi} }\sqrt{\epsilon_s} 
\simeq {1 \over  2 (2 \pi)^{3/2}} \sqrt{m_H}
\rightarrow m_H\simeq (m_H-0.841 \sigma_H)_{PDG},
\label{NWA2}
\ee
which suggests that a relationship between couplings involving $W+B$, $H+\phi$, and $Z+\phi$ mixing using the
customary value of the photon standard state energy \emph{inherits} a normalization choice for a massive $H$. 
This also relates the photon normalization to that of a massless mode generated \emph{from} that $H$.

Three additional coincidences will constrain either $m_W$ \emph{or} $m_H$:
\be
\alpha W1: \: {1 \over \sqrt{\pi}} g_W \:\sqrt{M_{Awb}}\simeq 1
\rightarrow m_W\simeq (m_W+0.356 \sigma_W)_{PDG},
\label{A1}
\ee
\be
\alpha W2: \: {1 \over 2\sqrt{4\pi}} (g_W \sin \theta_{H\phi}) \:M_{Awb} \simeq 1
\rightarrow m_W\simeq (m_W+0.407 \sigma_W)_{PDG},
\label{A2}
\ee
which both assign a standard state energy of the invariant $M_{Awb}$ for electromagnetic quanta, and
\be
\alpha W3: \: \sqrt{{\pi \over 2}} (g_W \cos \theta_{W\phi}) \:\sqrt{\mu_B} \simeq 1 \rightarrow
\begin{array}{l}
m_W\simeq (m_W+0.617 \sigma_W)_{PDG} \textnormal{ or}\\
m_H \simeq (m_H-0.283 \sigma_H)_{PDG},
\end{array}
\label{A3}
\ee
which assigns a standard state energy of $\mu_B$ for electromagnetic quanta resulting from $W+\phi$
mixing.  

The following coincidence determines only $m_W$:
\be
\alpha W13: \: {4 \pi g_B \sin \theta_{H\phi}\over \sqrt{2}} \simeq \sqrt{{<g_B \phi> \over \epsilon_{Awb}}}
\rightarrow m_W \simeq (m_W+0.55 \sigma_W)_{PDG},
\label{A13}
\ee
suggestive of mixing of $H+\phi$ to generate a massless mode with energy $\epsilon_{Awb}$. 
Two additional coincidences involve generation of a massless $H+\phi$ mode with charge quantization ${e_G \over 3}$:
\be
\alpha W41: \: {\sqrt{4 \pi}g_B \sin \theta_{W\phi}\over 3 \sqrt{2}} \sqrt{m_H \over \epsilon_{Yh\phi}}\simeq
{1 \over (2 \pi)^{2/3}}\rightarrow
\begin{array}{l}
m_W\simeq (m_W-0.04 \sigma_W)_{PDG} \textnormal{ or}\\
m_H \simeq (m_H+0.01 \sigma_H)_{PDG},
\end{array}
\label{A41}
\ee
suggestive of a state labeled by $m_H$ off mass shell towards $m_W$ generating a massless mode,
and
\be
\alpha A45: \: {4 \pi g_W \sin \theta_{H\phi} \over 3} \simeq \sqrt{\half}\sqrt{\mu_B \over \epsilon_{Yh\phi}}\rightarrow
\begin{array}{l}
m_W\simeq (m_W-0.004 \sigma_W)_{CDF} \textnormal{ or}\\
m_H \simeq (m_H-0.004 \sigma_H),
\end{array}
\label{A45}
\ee
suggestive of a state labeled by B generating a massless $H+\phi$ mode.
The plausible physical relevance of these various coincidences are not considered to be equivalent.

Exemplars pairing a kinematic coincidence with a photon normalization include\\
(\ref{KC1})+(\ref{A1}) $\rightarrow  m_W\simeq (m_W+0.356 \sigma_W)_{PDG}$ \emph{and}
$m_H\simeq (m_H+0.048 \sigma_H)_{PDG}$;\\
(\ref{KC3})+(\ref{NWA1}) $\rightarrow  m_W\simeq (m_W+0.159 \sigma_W)_{PDG}$ \emph{and}
$m_H\simeq (m_H-0.845 \sigma_H)_{PDG}$;\\
(\ref{KC3})+(\ref{A1}) $\rightarrow  m_W\simeq (m_W+0.356 \sigma_W)_{PDG}$ \emph{and}
$m_H\simeq (m_H-0.587 \sigma_H)_{PDG}$.

Alternatively, coincidences involving the reported renormalized value for the fine structure
constant ${1 \over \alpha_R (m_Z)}\simeq 127.951 $\cite{PDG2022,Schwartz} can be explored. 
The form of the running fine structure constant at this fixed point is taken to be
\be
\alpha_Z (\mu)={\alpha_R (m_Z) \over 1-{2 \alpha_R (m_Z) \over 3 \pi} g(\mu) \log {\mu^2 \over m_Z^2}},
\ee
where $g(\mu)$ sums over the squares of charges (in units of $e_G$) below charged pair productions at
momentum transfer $\mu$.  A few example coincidences follow:
\be
\alpha_Z2: \: {1 \over\sqrt{2}} {\sqrt{\alpha_Z(m_W)}\over 3 \sin \theta_{WB}} \simeq
\sqrt{4 \pi} {\epsilon_s \over m_W} \rightarrow
m_W\simeq (m_W-0.57 \sigma_W)_{PDG};
\label{ZA2}
\ee
\be
\alpha_Zr2: \: {4 \pi \sqrt{\alpha_Z(<g_B \phi>)}\over \sin \theta_{WB}} \sin \theta_{H\phi}
\sqrt{<g_B \phi> \over \epsilon_{Yh\phi}} \simeq 1 \rightarrow
\begin{array}{l}
m_W\simeq (m_W+0.19 \sigma_W)_{PDG} \\
\textnormal{or }m_H \simeq (m_H-0.07 \sigma_H);
\end{array}
\label{ZAr2}
\ee
\be
\alpha_Zr3: \: {1 \over 2 (2 \pi)^{3/2}}{\sqrt{4 \pi \alpha_Z(\epsilon_{Yhb})}\over \cos \theta_{WB}} \sin \theta_{W\phi}
\simeq {\epsilon_s \over \epsilon_{Yhb}} \rightarrow
\begin{array}{l}
m_W\simeq (m_W+1.31 \sigma_W)_{PDG} \\
\textnormal{or }m_H \simeq (m_H-0.04 \sigma_H);
\end{array}
\label{ZAr3}
\ee
\be
\alpha_Zr25: \: {\sqrt{4 \pi \alpha_Z(M_{Awb})}\over 3\sin \theta_{WB}} \cos \theta_{H\phi}
{\epsilon_{Yhb} \over M_{Awb}} \simeq {1 \over (2 \pi)^{3/2}} \rightarrow
\begin{array}{l}
m_W\simeq (m_W+0.09 \sigma_W)_{CDF} \\
\textnormal{or }m_H \simeq (m_H+0.01 \sigma_H).
\end{array}
\label{ZAr25}
\ee

Next, selected predictions involving possible combinations of kinematic coincidences (\ref{KC1})-(\ref{KC7})
with photon normalization coincidences (\ref{NWA1})-(\ref{A3}) will be presented
(recall ${1\over \alpha_{PDG}}\simeq 137.03599907$):
\begin{itemize}
\item (\ref{KC3})+(\ref{KC5})+(\ref{NWA1}) $\rightarrow  m_Z\simeq (m_Z+0.095 \sigma_Z)_{PDG},
m_W\simeq (m_W+0.173 \sigma_W)_{PDG}$ \emph{and}
$m_H\simeq (m_H-0.844 \sigma_H)_{PDG}$, or alternatively
${1 \over \alpha}\simeq 137.0357,
m_W\simeq (m_W+0.159 \sigma_W)_{PDG}$ \emph{and}
$m_H\simeq (m_H-0.846 \sigma_H)_{PDG}$;
\item (\ref{KC3})+(\ref{KC5})+(\ref{NWA2}) $\rightarrow  m_Z\simeq (m_Z+0.634 \sigma_Z)_{PDG},
m_W\simeq (m_W+0.257 \sigma_W)_{PDG}$ \emph{and}
$m_H\simeq (m_H-0.835 \sigma_H)_{PDG}$, or alternatively
${1 \over \alpha}\simeq 137.038,
m_W\simeq (m_W+0.159 \sigma_W)_{PDG}$ \emph{and}
$m_H\simeq (m_H-0.846 \sigma_H)_{PDG}$;
\item (\ref{KC7})+(\ref{KC11})+(\ref{A41}) $\rightarrow
{1 \over \alpha}\simeq 137.0329,
m_W\simeq (m_W+1.33 \sigma_W)_{PDG}$ \emph{and}
$m_H\simeq (m_H+0.47 \sigma_H)$.
\end{itemize}

A few observations are noteworthy.  First, the originally reported values of the kinematic angles
$\zeta W \simeq {\theta_{WB}\over 2}\simeq -\zeta Z$ for $W+B \rightarrow Z$ mixing were just
\emph{extremely} close numerical coincidences\cite{JLtoIJTP2022}. 
Subsequently, the analytic result (\ref{WZantimoving}) was found to \emph{require} this
to be the case, providing substantial motivational support for this approach.  Furthermore, there are
some cautionary observations that should be noted.  Previous coincidences involving (\ref{KC3}) and
(\ref{KC5}) using 2018 Particle Data Group (PDG) masses were very close, while they
are only somewhat close using the more recent 2022 PDG data.  Finally, one might be tempted to
utilize two compatible photon normalization coincidences with two massive kinematic coincidences to
attempt to calculate \emph{all} mass and coupling values.  For instance, if the four equations 
(\ref{KC3})+(\ref{KC5})+(\ref{NWA1})+(\ref{NWA2}) were all physically relevant, one could
\emph{calculate} all electroweak parameters as satisfying ${1 \over \alpha}\simeq 137.0368
(vs. 137.03599907),
 m_Z\simeq (m_Z+0.365 \sigma_Z)_{PDG},
m_W\simeq (m_W+0.215 \sigma_W)_{PDG}$ and
$m_H\simeq (m_H-0.84 \sigma_H)_{PDG}$. 
However, such a set of mathematically consistent relationships would make a disturbing inference about
the `arbitrariness' of the choice of the standard state photon energy $\epsilon_s$ in the photon
normalization terms, somehow mathematically fixing \emph{choice} of units of measurement.

From the representative set of kinematic coincidences presented,
it seems clear that PDG values provide far more kinematic coincidences than CDF values. 
It should be noted that there are other kinematic coincidences with a broad range of
\emph{subjective} plausibility that have been examined, but excluded for brevity. 
Some are consistent with the overall predictions described in this section, while others are incompatible.  

\subsection{Modeling using only anti-moving frames}

As has been mentioned, degenerate neutral $\Gamma=1$ causal bosons $W+B$ that can mix to form a massless
state also \emph{necessarily} generates a massive $m_Z = {m_W \over \cos \theta_{WB}}$ that moves in the
opposing Lorentz frame $\underline{\beta}_Z =-\underline{\beta}_W$.  Since the primary purpose of this
paper is to demonstrate a plausible pathway towards enhancing predictability and understanding of
standard model phenomenology, a final effort will be made to demonstrably model electro-weak phenomenology
consistent with kinematic coincidences involving kinematic angles $\zeta = \pm {\theta_{WB} \over 2}$
resulting from degenerate boson mixing.  The focus will remain only on the kinematics of \emph{known}
particles, exclusive of any undiscovered potential massive dark particles.

The phenomenological coincidences involving anti-moving frames that have been presented include
KC3 (\ref{KC3}), KC7 (\ref{KC7}), and KC11 (\ref{KC11}).  However, since KC11 involves energy scales
in the Lagrangian (not mass scales of particle states), the focus will be on particle state mixing to generate
new particles.  For convenience, the coincidences are repeated below, and plausible kinematic diagrams
for unitary mixing are exhibited in Figure \ref{WHWBtoZ}:
\be
KC3: p_W(\zeta W ^{m_Z}_{\mu_H}) \simeq p_H(\zeta W^{m_Z}_{\mu_B}),  \quad
KC7: \zeta H^{m_H}_{\mu_B}-\theta_{HB}\simeq -{\theta_{WB}\over 2}.
\ee
\onefigure{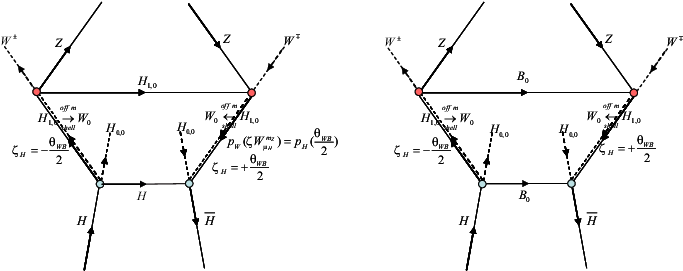}{Unitary diagrammatic representations of KC3 (left) and KC7 (right) generated by anti-moving
kinematics.}
The modeling should produce all known particle states in a kinematically consistent manner. 
For KC3, an {\bf{H}} and anti-{\bf{H}} colliding via exchange of an {\bf{H}} scatter into precisely
the needed reference frame $\zeta_H = {\theta_{WB} \over 2}$
(leaving the scalar H component unaffected), then go off-shell
as intermediate {\bf{W}}s with the same momenta.  The neutral
component of an intermediate {\bf{W}} has the momentum needed to mix with an orthogonal vector {\bf{H}}
to generate a {\bf{Z}}, while the charged components {\bf{W}}$^\pm$ which are not degenerate remain unaffected.
Thus the diagram generates all observed states.
For KC7, an {\bf{H}} and anti-{\bf{H}} colliding via exchange of a vector {\bf{B}} scatter into precisely
the needed reference frames $\zeta_H = \pm{\theta_{WB} \over 2}$
(leaving the scalar H component unaffected), then go off-shell
as an intermediate {\bf{W}} with the correct kinematic parameter needed to mix with a {\bf{B}} to generate
a {\bf{Z}}.  The charged components {\bf{W}}$^\pm$ which are not degenerate remain unaffected, likewise
generating all observed states.  This diagram is suggestive of the intermediate {\bf{B}}s as off mass-shell
tachyonic {\bf{H}}s that can unitarily mix to generate {\bf{Z}} and {\bf{A}} bosons.

\setcounter{equation}{0}
\section{Discussion and Conclusions}
\indent

The model described in this paper (examining bosons) and its companion part I
paper (which examined fundamental fermions) has been demonstrated to mirror causal spinor
fields that exhibit properties consistent with observed leptons and bosons.  
Just as is the case for Dirac fermions, the $\Gamma=1$ boson representation
spinors have the kinematically dependent forms that allow the equation of motion to be linear
in the energy-momentum \emph{operators}, simplifying cluster decomposability.  However, unlike the
fermion representation, there are a set of degenerate (massive and massless) spinor fields corresponding to
a vanishing eigenvalue for the equation of motion, inclusive of both vector and scalar components. 
This allows the degenerate set of spinors to be unitarily
mixed to generate physical states that continue to satisfy the defining equation of motion, suggestive of a
meaningful application of the formulation to describe electro-weak bosons.

Unitary mixing of orthogonal spinors consistent with one resultant state
possibly being massless defines analytic forms
relating the kinematic angles of the involved spinors in terms of the invariant mass of that resultant.  In particular,
the unique relationship of an {\bf{H}} spinor with an orthogonal $\mathbf{\Phi}$
spinor with (tachyonic) mass $\mu_\phi={m_H \over \sqrt{2}}$
to generate a resultant spinor of mass $m_H$, might \emph{also} generate an
orthogonal massless (Goldstone-like) mode
utilizing the same invariant (extended Poincare) group label $m_H$.  Furthermore, mixing $W+B$ at invariant energy
$m_Z={m_W \over \cos \theta_{WB}}$ analytically \emph{requires} that the
kinematic frame of the {\bf{W}} must be equal but opposite that of the resultant {\bf{Z}}, with a value precisely
half the unitary mixing angle ${\theta_{WB} \over 2}$.  These (and other) analytic relationships compelled the author
to search for meaningful kinematic coincidences in a manner analogous to the examination of planetary motions
by Kepler and others in hopes of discovering a set of rules that can lead to the development of acceptable
supplementations to standard model canon.

Several representative (but by no means comprehensive) analytic relationships and kinematic coincidences that fall within a
(subjectively) chosen range of plausibility and consistency with measured electro-weak parameters
have been presented for demonstrative purposes.  The selection includes exemplars that
illustrate how the formulation might provide insight into how the Higgs sector
provides kinematic structure to the \emph{massive} W and Z spinor fields,
as well as substantive support for massless modes.
Furthermore, it should be noted that the formulation has the potential to provide
insight into states that are electromagnetically dark, yet have physical and cosmological significance.

To conclude, an attempt has been made to develop exemplary model candidates consistent with what was initially
an extremely close experimental kinematic coincidence involving $W+B\rightarrow Z$ mixing that has now
been demonstrated to be an analytic necessity.  The exemplars predict kinematic parameters consistent with
present data. 
The approach recognizes that the development of a \emph{compelling} model ultimately embraces
the most elegant combination of analytic relationships with plausible kinematic coincidences that
remain consistent with future measurements, as well as predictive of yet to be discovered phenomena.


\end{document}